\documentclass[fleqn,twoside]{article}
\usepackage{espcrc2}

\usepackage{epsfig,amssymb}
\usepackage[figuresright]{rotating}

     \newcommand{\pathnow}{}
\font\twelvegoth=eufm10 at 14.4pt
\newfam\gothfam
\textfont\gothfam=\twelvegoth

\font\twelveron=eusm10 at 14.4pt
\newfam\ronfam
\textfont\ronfam=\twelveron

\newcommand{\nc}{\newcommand}
\nc{\rf}[1]{Fig.\,\ref{#1}}

\nc{\eps}{\varepsilon}
\nc{\la}{\lambda}
\nc{\ga}{\gamma}
\nc{\Ga}{\Gamma}
\nc{\de}{\delta}
\nc{\De}{\Delta}
\nc{\al}{\alpha}
\nc{\be}{\beta}
\nc{\ra}{\rightarrow}
\nc{\llra}{\longleftrightarrow}
\nc{\Llra}{\Longleftrightarrow}
\nc{\Ra}{\Rightarrow}
\nc{\nf}{\infty}
\nc{\Lra}{\Longrightarrow}
\nc{\beq}{\begin{equation}}
\nc{\eeq}{\end{equation}}
\nc{\beqa}{\begin{eqnarray}}
\nc{\eeqa}{\end{eqnarray}}
 \nc{\beql}[1]{\begin{equation}\label{#1}}
\nc{\req}[1]{Eq.\,(\ref{#1})} 
\nc{\bfi}{\begin{figure}}
\nc{\efi}{\end{figure}}
\nc{\prh}[2]{\left(\hspace{-0.2cm}\begin{array}{c}#1 \\#2 \end{array}\hspace{-0.2cm}\right)}
\nc{\prf}[3]{\left(\hspace{-0.2cm}\begin{array}{c}#1 \\#2 \end{array}\hspace{-0.1cm};#3\right)} 
\nc{\cro}[2]{\left[\begin{array}{c}#1 \\#2\end{array}\right]}
\nc{\fx}[5]{{_{#1}F_{#2}}\prf{#3}{#4}{#5}}
\nc{\ff}[4]{{_{#1}F_{#2}}\prh{#3}{#4}}
\nc{\bla}{
}  
\nc{\blu}{
}  
\nc{\gre}{
}  
\nc{\greh}{
}  
\nc{\gryl}{
} 
\nc{\gry}{
} 
\nc{\gryh}{
} 
\nc{\perh}{
}  
\nc{\per}{
}  
\nc{\red}{
}  
\nc{\vio}{
}  
\nc{\yel}{
}  
\nc{\ora}{
}  
\nc{\whi}{
}  
\nc{\blui}{
}  
\nc{\redp}{
}  
\nc{\viob}{
}  
\nc{\redi}{
}  
\nc{\redx}{
}  
\nc{\yelh}{
}  

\nc{\ova}[3]{\begin{picture}(1000,50)(0,0)\put(#1,20){\thicklines\oval({#2},40)}\put(10,15){#3}\end{picture}}

\nc{\clbox}[1]{
\begin{center}\framebox{#1}\end{center}
}
\nc{\flbox}[1]{\fboxrule=5pt\framebox{#1}\fboxrule=2pt}

\newcommand{\AmS}{{\protect\the\textfont2
  A\kern-.1667em\lower.5ex\hbox{M}\kern-.125emS}}

\title{Strangeness and thresholds 
       of phase changes\\ in relativistic heavy ion collisions }

\author{Johann Rafelski\address[DP]{Department of Physics, University of Arizona, Tucson, AZ 85718, USA}\thanks{JR thanks the Centre for the Subatomic Structure of Matter (CSSM) Adelaide 
 and Prof. A. G. Williams for kind hospitality in 
Australia,  which made this contribution possible. Work supported in part by a 
grant from: the U.S. Department of
Energy  DE-FG02-04ER4131. E.mail: Rafelski@Physics.Arizona.EDU}
      and
 Jean Letessier\address[LPTHE]{Laboratoire de Physique Th\'eorique et Hautes Energies\\
Universit\'e Paris 7, 2 place Jussieu, 75251 Cedex 05, France}\thanks{LPTHE, Univ.\,Paris 6 et 7 is: Unit\'e mixte de Recherche du CNRS, UMR7589. E.mail: JLetes@LP\-THE.Jussieu.FR. 
}
}
       
\begin{document}

\begin{abstract}
We discuss how the dynamics of the evolving hot fireball of 
quark--gluon matter impacts   phase transition between 
the deconfined and confined  state of matter. The rapid expansion 
of the fireball of deconfined matter created in heavy ion collisions
facilitates formation of an over-saturated strange quark 
phase space. The related excess abundance of strangeness is compensating 
the suppression of this semi-heavy quark yield
by its quark mass. In addition, the dynamical
expansion of colored quanta pushes against the vacuum 
structure, with a resulting supercooling of the transition
temperature. We address the status of the  search for the 
phase boundary as function of reaction energy and collision 
centrality and show evidence for a change in reaction mechanism at
sufficiently low energies. The phase diagram derived from the 
study of hadron production conditions shows two boundaries, one corresponding to
the expected transition between confined and deconfined matter, with 
a downward temperature shift, and the other a high quark density hadronization
which appears to involve heavy effective quarks, at relatively large temperatures. 
 
\vspace{1pc}
\end{abstract}

\maketitle

\section{INTRODUCTION}\label{intro}
Quark--gluon plasma phase
(QGP) is the {\it equilibrium} state of  deconfined 
hadronic matter at high temperature
and/or density.  It is   believed that  this state has  been 
present in the early Universe, 10---20$\mu$s into 
its evolution. Today, we recreate the conditions of the 
early Universe in laboratory experiments colliding atomic
nuclei (heavy ions) at highest available energies.
Colliding heaviest nuclei ,
we explore  a domain of space and time  
much larger than normal hadron  size, 
in which color-charged  quarks and gluons are
propagating  constrained by external 
`frozen  vacuum', which abhors color. For the past
four years, dedicated experimental program has been
carried out at the RHIC collider~\cite{RBRC}.

It is an open  question,  if  within the short
time, $10^{-22}$--$10^{-23}$~s,  available in a laboratory heavy ion
collision  experiment, the confined color frozen nuclear phase 
can melt and turn into the deconfined QGP state of matter. There is
no   answer  available today, nor as it seems,
will a first principles simulation of the dynamic heavy ion
environment become available in the foreseeable   future.  
Our attention  turns to the study of QGP observables.

\begin{figure}[htb]
\vskip 0.2cm
\centerline{
\psfig{width=7cm,clip=,figure=\pathnow 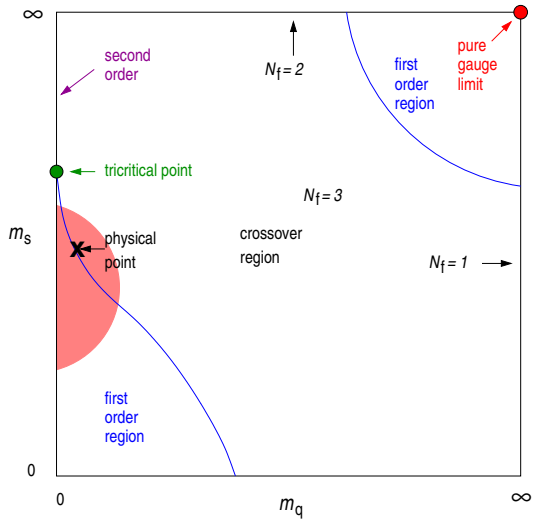}
}\vspace*{-0.7cm}
\caption{
\label{2+1Phases}  
Phases of QCD thermal phases in the $m_q$-$m_s$ plane, 
adapted from Ref.\,\cite{Peikert:1998jz}.
}
\end{figure}

The pattern of production  of  strangeness $s$, and more generally,
strange hadrons, is an important observable  of QGP; for
further theoretical details and historical developments
see our book~\cite{CUP}. Here, we will discuss how  strangeness production
within  a   fireball of rapidly expanding matter
can   strengthen  the phase transition between QGP and HG.
This is of  importance since the QCD thermal state
turns out to remain below a first order transition for a 
physical set of masses of two low mass $m_q/T \to 0$, and one 
semi-heavy quark of mass $m_s/T\simeq 1$.
This is illustrated in the $m_s$--$m_q$ plane 
in  Fig.~\ref{2+1Phases}~\cite{Peikert:1998jz}.

 The presence of  a positive quark (baryo) chemical potential
$\mu_{\rm B}\equiv 3\mu_q$
enhances  the quark number, and suppresses anti-quark 
number. Still, the   effect of $\mu_{\rm B}$  is to increase the pressure,
and the effect is important when $T\simeq \mu_q$.  
The nearly massless quarks respond more strongly to the finite 
baryo-chemical potential, than do massive hadrons, and that is why
the quark-phase response matters more. For this reason, a
1st-order phase transition is expected to arise at a finite value
of the chemical potential~\cite{Fodor:2004nz,Karsch:2003jg,Bernard:2004je}.
Baryon density, as expressed by the value of the baryo-chemical potential
$\mu_{\rm B}$, is relatively low at RHIC, where $\mu_{\rm B}\simeq 25$ MeV.
At the much higher energy at LHC,  the expected value of baryo-chemical
potential is of magnitude $\mu_{\rm B}\simeq 1$--$ 2$ MeV~\cite{LHCPred}.

We recognize in Fig.~\ref{2+1Phases} that if we could compensate
the  effect of the finite strange quark 
mass, this would suffice to  move the 
physical point to phase transition region at vanishing
baryo-chemical potential. In chemical non-equilibrium,
the Fermi distribution function takes the form~\cite{cool}: 
\begin{equation} \label{dists}
{d^6N_s\over d^3p\,d^3x} = {\gamma_se^{-\sqrt{m_s^2+p^2}/T} 
                 \over 1+\gamma_se^{-\sqrt{m_s^2+p^2}/T} } ,
\end{equation}
We recognize that for $\gamma_s >1$ there is compensation of the finite 
mass effect~\cite{Letessier:2005qe}. In the limit  $\gamma_s \to e^{m_s/T}$
the strange quark would play a similar role as a light quark
and we  expect a rather strong first order phase transition 
even at vanishing baryo chemical potential.

The question we address next is, if such an over-saturation
of the strangeness phase space near to the phase transition
is possible in the dynamical environment we study. We also 
look at any  further dynamical effects associated
with the explosive flow of deconfined matter, and search 
to understand if this   assists   development of the singular
phase behavior.  Following these studies we
explore the current RHIC and future LHC environments from
this perspective and determine the phase limit between
hadron and quark phases. Our study of the experimental 
hadronization conditions suggests further 
that  at relatively low collision energies the final state hadrons 
emerge  from a high baryon density phase
with the mass  of constituents  being indicating   
constituent quark mass  phase~\cite{Roizen:2004cy}.

\section{QGP PHASE OVER\-SATURATION}
\label{Tphase}
The   fireball of QGP created in heavy ion collisions is initially 
significantly more dense and hot than after its expansion towards 
final breakup condition. This   expansion dilutes the high strangeness 
yield attained in the  initially very dense and hot phase. Contrary 
to intuition, this can result in an over-saturation of the chemical
abundance, even if the initial state is practically strangeness free. 

In order  to understand this effect qualitatively, 
consider the yield of strange quark pairs 
 in Boltzmann approximation, at a temperature $T=T_e$, time $t=t_e$,
within the volume $V=V_e$ we have:
\begin{equation}
N_s(t_e)= \gamma_e {2 V_eT_e^3\over \pi^2}
            x_e^2 K_2(x_e), \   x_e={m_s(T_e)\over T_e}.
\end{equation}
We have shown above that the mass of strange quark is     $T$-dependent
as the scale of energy at which its value is determined is in the domain
where a rapid mass change occurs~\cite{CUP}. Moreover, as we shall argue at 
the end of this work, there is evidence for presence of heavy quasi-partons
in the deconfined phase, and such an effective thermal mass is strongly temperature
dependent. 

We choose  the  value of  $T_e$ to be the point 
where the system has nearly reached chemical equilibrium abundance
in QGP, with $\gamma_e^{\rm QGP}=1$ (superscript QGP reminds us that
we are considering the deconfined phase).  
 We  assume that the continued expansion  
 preserves entropy as is appropriate for an ideal 
liquid. Since the entropy is governed by essentially massless 
quark--gluon quanta, this implies that $VT^3=$ Const.\,.  
Model calculations show that, for typical values of $T_e$,  the change 
in the absolute number of strange quark pairs  has  
essentially stopped for $T<T_e$. Then for $t_1>t_e$:
\begin{equation}
{N_s(t=t_1)\over N_s(t=t_e)}\simeq 1 =
   \gamma_s(T_1)  {x_1^2K_2(x_1)\over x_e^2K_2(x_e)}.
\end{equation}
Since the function $x^2K_2(x)$ is a monotonically falling function
(see Fig.\,10.1 p197 in Ref.~\cite{CUP}),  
in general  $\gamma_s^{\rm QGP}(t_1)>1$. In the non-relativistic limit 
$m>T$, that is $x>1$:
\begin{equation}
\gamma_s^{\rm QGP}(T_1) = 
{  x_0^2K_2(x_0)  \over x_1^2K_2(x_1)}  > 1 .
\end{equation}

How large can $\gamma_s(T_1)$ be? Should the QGP breakup occur
from a supercooled state with $T_1=140$ MeV, than it is appropriate
to consider $m_s(T_1)/m_s(T_0)\simeq 1.5$--2, with  $T_0\simeq 210$ MeV. 
To convert from temperature dependence to the energy scale dependence, 
we set $\mu\simeq 2\pi T$
and thus for $m(T=T_0)$, we need and estimate of $m_s(\mu=1.3\,{\rm GeV})$. 
The PDG value is $m_s(\mu=2\,{\rm GeV})= 80$--130 MeV. Thus 
 assuming $m_s(T_0)\simeq 140$ we find
$\gamma_s^{\rm QGP}(T_1)\to 1.9$ . As this example shows it is possible to 
 nearly completely compensates
the effect of the  strange quark mass, see Eq.\,(\ref{dists}).

Though $\gamma_s$ is a useful tool to understand how strangeness
can help to facilitate phase transition, it is important to
remember that it is merely a  quantity, which relates actual abundance 
of strangeness to the equilibrium abundance,  
\begin{equation}
n_s\simeq \gamma_s n_s^\infty, \quad  n_s^\infty\equiv n_s(\gamma_s=1).
\end{equation}
A more direct    way to see how 2+1 flavors turns into a  3-flavor QCD
is to note that we hope and expect that the dynamics of fireball expansion
could help us to reach the condition: 
\begin{equation}
n_s(T_1)\simeq n_u(T_1)\simeq n_d(T_1)
\end{equation}
even if at point of chemical equilibrium, 
$n_s(T_e)\simeq 0.5 n_i(T_e), i=u,d$. 

In physical terms relative importance of strangeness increases
since strangeness yield is not reduced along with light quark and 
gluon yields during the dense matter expansion. One can argue this in 
several ways. One is to look at the buildup of collective expansion of the 
fireball of matter which requires conversion of thermal energy into  kinetic energy of 
collective  motion.   The   entropy density decreases, but the expansion
assures that the total entropy remains constant, or slightly  increases. 
As the energy is transfered into the transverse expansion, strange quark pair yield, 
being  weaker coupled, remain least influenced by this loss which mainly consumes
light quarks and gluons.  

We thus conclude that  in the event the initial
 conditions present in QGP are 
sufficiently extreme to generate  strangeness rapidly
and abundantly, one can expect 
over-population of strangeness in the 
final QGP breakup condition can   facilitate
the  occurrence of a  phase transition at small and vanishing
baryon density. In fact there is strong evidence that 
this can happen: we have performed model
calculations of strangeness (over)population 
at RHIC  prior to the first experimental results 
becoming available~\cite{Rafelski:1999gq}.
We found    $\gamma_s(T_1)^{\rm QGP}<1.2$ .

We believe that greater values of $\gamma_s^{\rm QGP}$  
can be easily arrived at: in the model considered the 
instantaneous establishment of transverse expansion 
speed cut into the lifespan of the QGP phase and thus 
reduced the production yield of strangeness.
 In addition the
quark mass chosen was significantly above the currently 
preferred value ($m_s=105\pm 25 $MeV) which has 
further cut into the rate of production of strangeness.

However, some other studies have 
found considerably smaller values of  $\gamma_s^{\rm QGP}$ 
at RHIC~\cite{Biro:1993qt,Pal:2001fz,He:2004df}. In part, this is due 
to employment of dynamical QGP equilibration models that do not allow
gluon chemical equilibration. This maybe  due to lack in these approach 
of rapid gluon equilibration  processes, which are presently not fully 
understood.  We assumed rather rapid glue chemical equilibration
with relaxation time shorter than 1.5 fm. 
Moreover,  we note somewhat unrealistically low values used for the 
coupling constant $\alpha_s$   noted --  we use QCD measured value:
$\alpha_s(M_Z)=0.118$ and use 2 and higher loop evolution to obtain 
the low energy scale values~\cite{impact}.  We believe that 
if not at RHIC, than at LHC, we should
expect at QGP hadronization a substantial phase space 
excess of strange quarks.

\section{EXPLOSIVE MATTER FLOW}
\label{exp}
Gibbs characterized  in detail the  boundary conditions  between 
two phases of matter. The Gibbs condition of importance here
is:
 \begin{equation}\label{gibbs}
P\equiv P_A-P_B=0\,.
\end{equation}
This condition assures force equilibrium: the boundary of 
phase $A$  does 
not move sine phase $B$ balances it exactly. This condition is
 thus  assuring the dynamical 
stability of the phase boundary between  phases $A$ and $B$.

The covariant characterization of the Gibbs condition
requires the introduction of the energy--momentum tensor
$T^{\mu \nu}$. In the laboratory rest frame its components are:
\begin{equation}\label{Tmunurest}
 \widehat T^{ij}=P\delta_{ij},\  
  \widehat T^{i0}= \widehat T^{0i}=0,\ 
  \widehat T^{00}=\varepsilon\,.
\end{equation}
where $\varepsilon$ is the energy density. The
Latin indices as usual refer to space component $i=1,2,3$
and the wide hat indicates the laboratory frame. 

Gibbs considered
a space-like  surface along which pressure difference had to vanish. This 
surface is invariantly characterized by a  normal four-vector: 
\begin{equation}\label{normal}
 n^\mu=(0,\vec n) .
 \end{equation}
We take as the covariant, frame of reference independent
 statement of the Gibbs condition Eq.\,(\ref{gibbs}):
\begin{equation}\label{gibbscov}
T^{\mu \nu}n_\mu n_\nu\equiv 
 T_A^{\mu \nu}n_\mu n_\nu - 
 T_B^{\mu \nu}n_\mu n_\nu =0 . 
 \end{equation}
 
We now consider matter subject to expansion flow.  
$\vec v$ is the velocity of the local
matter element, and its 4-velocity is:
\begin{equation}\label{4v}
u^\mu=(\gamma,\vec v \gamma), \quad  \gamma={1\over \sqrt{1-v^2}}.
\end{equation}
The natural presence of two Lorentz vectors  $n^\mu, u^\mu $, assures that 
we cannot transform away the effect of motion, the colored state 
pushes against Gibbs surface where the phase boundary is located. 
We recognize this as the hadronization hyper-surface, where the 
final state hadrons are born. 
 
The components of interest in
the energy momentum tensors are the pressure components:
\begin{equation}\label{Tijv}
 T^{ij}=P\delta_{ij}+(P+\varepsilon)\frac{v_iv_j}{1-\vec v^{\,2}}\,.
 \end{equation}
The balance of forces
between the deconfined  QGP and confined hadron phase 
 comprises the effect of the vacuum which confines color.
 This pressure, introduced  in 
bag models of hadrons for the first time, is
traditionally referred to as the bag constant, 
${\cal B}\simeq (0.2 {\rm GeV})^4$~\cite{Letessier:2003uj}.
This  vacuum structure can be represented within $ T^{ij}$ by:
\begin{equation}\label{vacpe}
P_V=-{\cal B},\quad P_V+\varepsilon_V=0 .
 \end{equation}
The   vacuum structure component is thus not entering  the 
dynamical flow term, the last term in Eq.\,(\ref{Tijv}).

We obtain from Eq.\,(\ref{gibbscov}):
\begin{equation}\label{gibbsexpl}
 {\cal B}=P_{\mbox{\scriptsize p}}+
      (P_{\mbox{\scriptsize p}}+\varepsilon_{\mbox{\scriptsize p}})
\frac{\kappa  v^2}
       {1-v^{2}}\,,
\quad
\kappa=\frac{(\vec v\cdot \vec n)^2}
              {v^2}\,.
 \end{equation}
Here,
\begin{equation}
P_{\mbox{\scriptsize p}}\equiv 
     P_{\mbox{\scriptsize p}}^{\mbox{\scriptsize QGP}}
   - P_{\mbox{\scriptsize p}}^{\mbox{\scriptsize HG}},\quad
\varepsilon_{\mbox{\scriptsize p}}^{\mbox{\scriptsize p}}\equiv 
     \varepsilon_{\mbox{\scriptsize p}}^{\mbox{\scriptsize QGP}}
   - \varepsilon_{\mbox{\scriptsize p}}^{\mbox{\scriptsize HG}}, 
 \end{equation}
are the {\it particle} pressure and energy density components in the 
pressure and energy density, from which  any  vacuum term has been separated. 
For $\vec v\to 0$, the conventional Gibbs condition reemerges:
\begin{equation}
P_{\mbox{\scriptsize p}}^{\mbox{\scriptsize HG}}
      =P_{\mbox{\scriptsize p}}^{\mbox{\scriptsize QGP}}- {\cal B}.
   \end{equation}

Eq.\,(\ref{gibbsexpl}) describes the pressure of motion of color charged matter against the 
vacuum structure which is pushed out as color cannot exist there\cite{Rafelski:2000by}; 
we can speak of color wind~\cite{Csorgo:2002kt}.
The magnitude of the   flow effect  on the value of temperature of the 
phase boundary  is relatively large. This is due to
a rather large, in comparison to the pressure, energy density.
In presence of a phase transition,  we can expect
that the discontinuity in particle energy is about 6--7 times greater than
the discontinuity in the particle pressure in  Eq.\,(\ref{gibbsexpl}) ---
we recall that only
the total pressure is continuous as expressed by the covariant Gibbs 
condition, Eq.\,(\ref{gibbscov}).

As a consequence,  we expect a significant down-shift in the critical temperature:
the moving colored fields cause  the quark matter to  super-cool by as much as 
$\Delta T=25$ MeV \cite{Rafelski:2000by}. This means that if the 
equilibrium phase crossover  were to occur at $T=165$ MeV, the dynamical 
cross over would be postponed to the value $T=140$ MeV. We do not know
at this time if the collective and rapid outflow of QCD can turn a  cross-over  into 
a phase-transition.
 
\section{STRANGENESS AND ENTROPY}
\label{StatHad}
The study of strangeness production requires  a model of 
particle production in heavy ion
collision. To describe the yields of particles produced we employ
the statistical hadronization model (SHM) as implemented in 
the SHARE program package~\cite{share}. SHM  is by definition a model of 
particle production in which the birth process of each particle 
fully saturates (maximizes) the quantum mechanical probability amplitude, and
thus, the relative  yields are determined solely 
by the appropriate integrals of the
accessible phase space.
 For a system subject to global dynamical 
evolution such as collective flow,  this is understood to apply 
within each local  co-moving frame. Empirical 
evidence shows that the full SHM is capable to describe the experimental
hadron particle data  rather precisely.

When particles are produced in hadronization, 
we speak of chemical freeze-out.  
In order to arrive at an adequately precise understanding of 
the entropy production contained in the final state in the 
overall hadronic particle yield, 
one has to be sure to include all the hadronic resonances
which decay feeding into the individual yield considered, {\it e.g.}, the decay 
 $K^*\to K+\pi$ feeds into $K$ and $\pi$ yields. It is important,
in this context, to note that the magnitude  of resonance
production  is sensitive to the (freeze-out) 
temperature at which these particles are formed. 

The indirect  resonance decay contribution, to particle 
yields, is dominant for the case of the  pion yield at all 
hadronization conditions considered in literature. This happens  
even though each resonance contributes relatively 
little in the final count. However,  
the large number of resonances which  contribute compensates 
and the sum of small contributions   competes with the 
direct pion yield. For the more heavy hadrons, generally
 there is a dominant  contribution from just a few particles.
 
In order to study the properties of the 
phase of matter that was created early on in the reaction, we  study  
strangeness and entropy  production. Entropy is an observable of similar character  
as strangeness, it is preserved or slightly increasing in the evolution of the
reaction system. In fact, it is a `deeper' observable than strangeness, 
which is produced by reactions occurring mostly after formation of entropy 
has occurred. Entropy enhancement, observed in terms of specific (per electrical charge) 
hadron multiplicity, has been perhaps  the first indication  of the
new physics reach of CERN-SPS experimental heavy ion program~\cite{entro}.

Hadron formation from QGP phase has  to absorb the high
entropy  content of QGP which originates in broken color bonds.
The lightest hadron is pion and most entropy per energy 
is consumed in hadronization  by producing these
particles abundantly. It is thus important to 
free the yield of these particles from the chemical 
equilibrium constraint. 

At RHIC, we have very rich data field with yields of many final state particles
available. This allows us to   consider both, the  yield of strangeness
 per (nearly conserved) entropy and per (exactly conserved) baryon number. 
In addition, 
we look at the cost in final state thermal energy to make a strange quark 
pair.

In the  QGP, the dominant entropy production 
occurs during  the initial glue thermalization, and the  thermal
strangeness production occurs in parallel and/or just a short time later~\cite{Alam:1994sc}.  
The entropy  production occurs predominantly 
early on in the collision during  the thermalization phase.  
Strangeness production by gluon fusion 
is  most effective in the early, high temperature environment,
however it continues to act during the evolution of
the   hot deconfined phase until hadronization \cite{RM82}.

Both strangeness and entropy are nearly conserved in
the evolution towards hadronization
and thus the final state hadronic yield analysis value for $s/S$ is closely
related to the thermal processes  in the fireball at $\tau\simeq 1$--2 fm/c. 
We believe that, for reactions in which the system approaches strangeness
equilibrium in the QGP phase, one can expect a prescribed ratio of 
strangeness per entropy, the value is basically the ratio 
of the QGP degrees of freedom.

We estimate the magnitude of $s/S$ deep in the QGP phase, considering  
the hot   stage of the reaction~\cite{Rafelski:2004dp}. For  an   equilibrated 
non-interacting QGP phase with perturbative properties:
\begin{eqnarray}\label{sdivS}
{s \over S}   &=& 
\frac{ (3/\pi^2) T^3 (m_{  s}/T)^2K_2(m_{  s}/T)}
  {(32\pi^2/ 45)  T^3 
    +n_{\rm f}[(7\pi^2/ 15) T^3 + \mu_q^2T]} , \nonumber \\
 &=& 
 {0.027\over {1+ 0.054 (\ln \lambda_q)^2} }\,.
\end{eqnarray} 
Here, we used for the number of flavors $n_{\rm f}=2.5$ and $m_{  s}/T=1$. We 
see that the result is a slowly changing function  of $\lambda_q$;
for large $\lambda_q\simeq 4$, we find at modest SPS energies, the 
value of $s/S$ is reduced by 10\%. Considering 
the slow dependence on $x=m_{  s}/T\simeq 1$ of $W(x)=x^2 K_2(x)$ there is 
minor dependence on the much less variable temperature $T$.

The dependence on the degree of chemical equilibration 
which dominates is easily obtained separating the different 
degrees of freedom, and for simplicity, we look here at
the case $\lambda_q\to 1$:
\beql{sdivS2}
{s \over S}=f(\alpha_s) 0.027 {\gamma_s  \over 
  {0.38 \gamma_{\rm G} + 
         0.12 \gamma_s +
         0.5\gamma_q   
}}
\,.
\eeq
All $\gamma_i$ refer here to QGP phase. $f(\alpha_s)$ is an unknown factor that 
accounts for the  interaction effects --- up to this point, we had assumed that 
these are canceling, with  $f(\alpha_s)\to 1$ . In any
case, seen the dependence of \req{sdivS2} on $\gamma_i$, we expect to see a gradual 
increase in $s/S$ as the QGP source of 
particles approaches chemical equilibrium with increasing 
collision energy   and/or  increasing volume. 

Since the  ratio $s/S$ is established early on in the 
reaction,    the above relations, and the associated 
chemical conditions  we considered, probe
 the early hot  phase of the fireball. In fact,
 at hadron freeze-out the 
QGP picture used above does not apply. Gluons are likely
to freeze faster than quarks and both are subject to much
more complex non-perturbative behavior. However, the value of
$s/S$ is nearly preserved from the hot QGP to the  final state
of the reaction.
 
How does this simple prediction compare to experiment?
Given the statistical parameters, we can evaluate the 
yields of particles not yet measured and obtain
the rapidity yields of  entropy, net baryon number, net strangeness,
and thermal energy, both for the total reaction
system and also  for the central rapidity condition, also
 as function of  centrality.

\begin{figure*}[!bt]
\centerline{
\psfig{width=7.cm,angle=0,clip=,figure=\pathnow  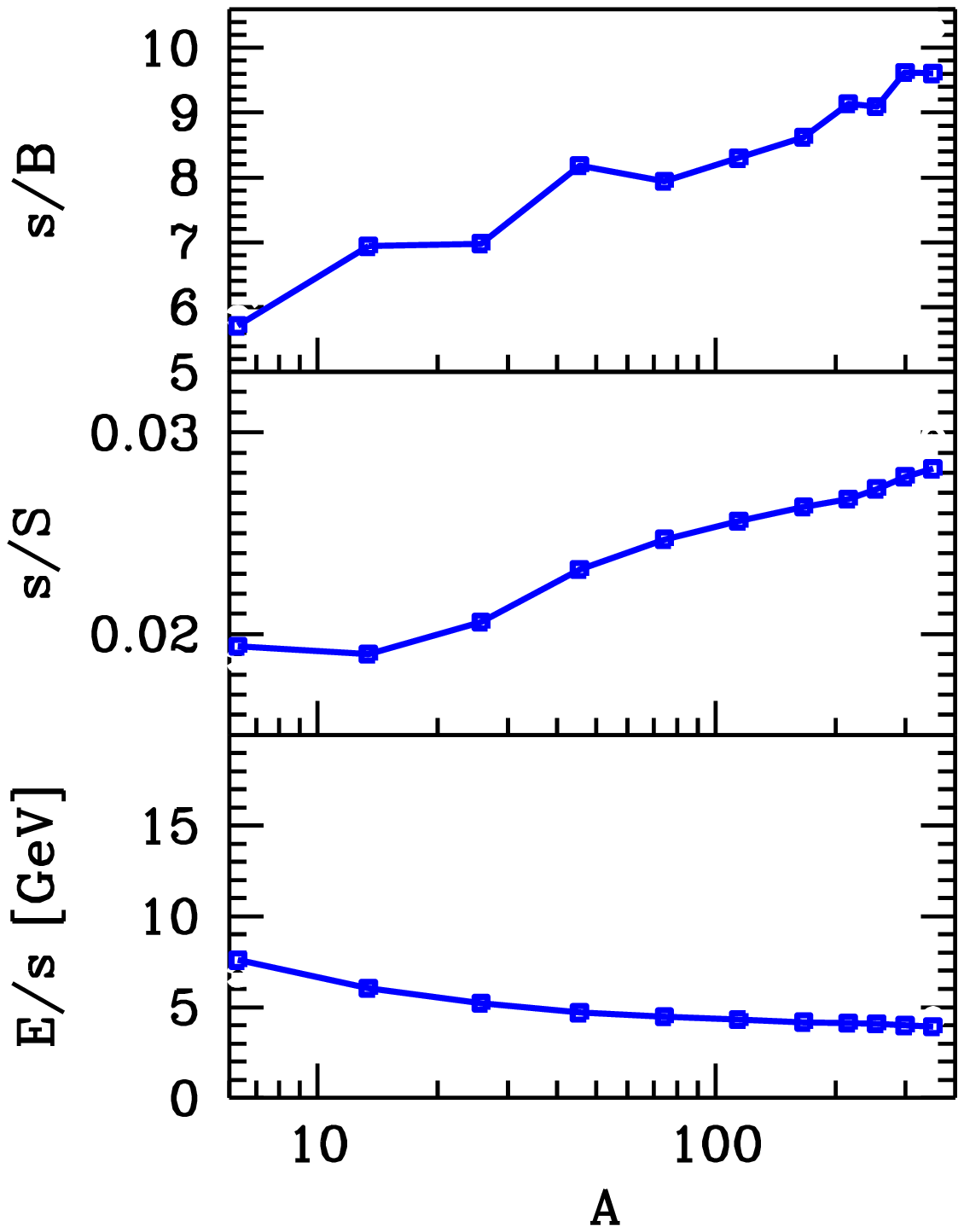}
\hspace*{.3cm}\psfig{width=7.cm,clip=,figure=\pathnow 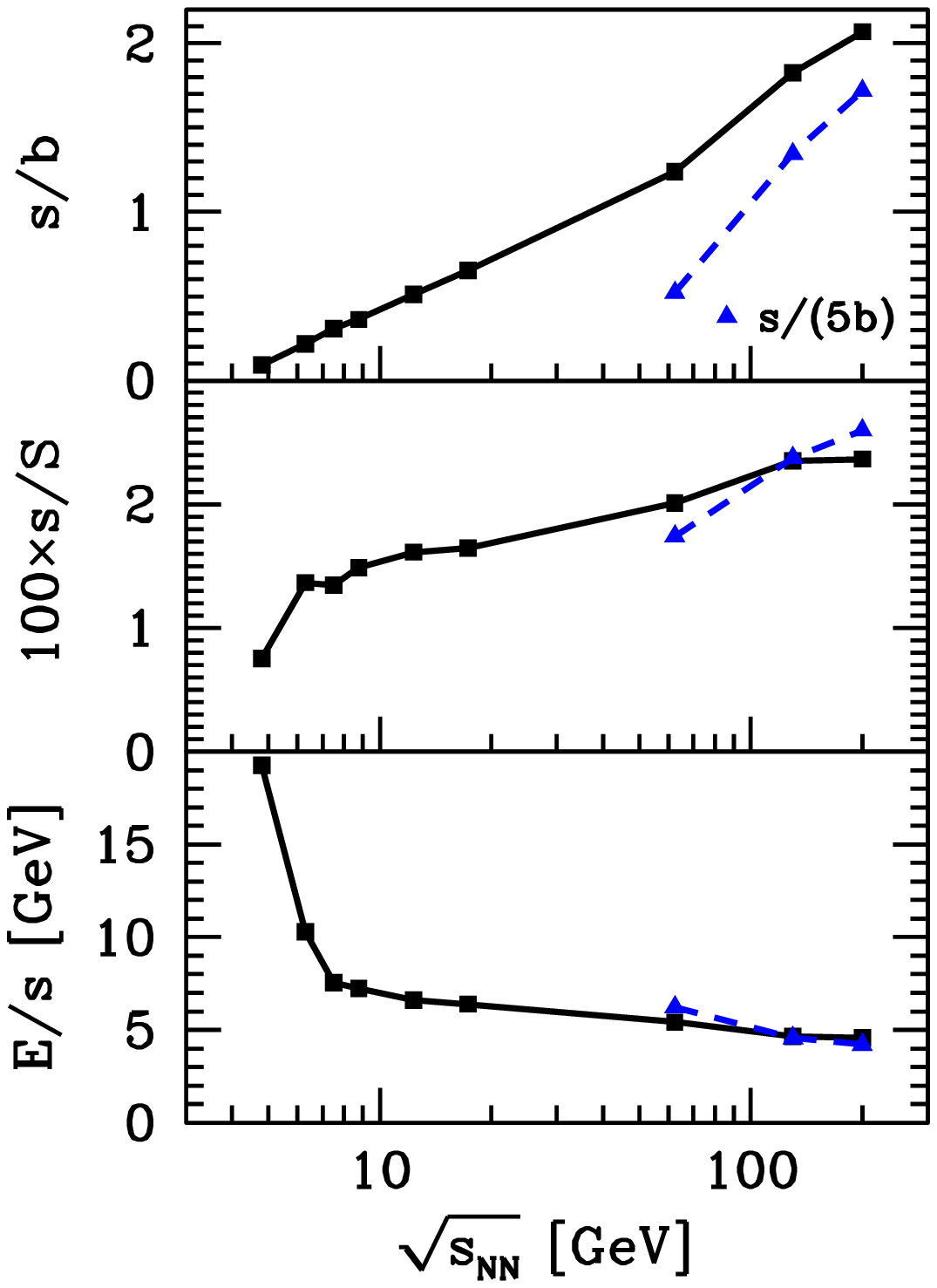} 
}
\vspace*{-1.cm}
\caption{\label{PEST}
Strangeness per net baryon $s/B$, 
strangeness per entropy $s/S$,   and   $E_{\rm th}/s$ the thermal energy  
cost to make strangeness.  Left:
as a function of centrality, Right as function of reaction energy,
adopted from \cite{Letessier:2005qe,Rafelski:2004dp}
The results on right 
include the central rapidity conditions at RHIC energies
(dashed, blue) lines. The actual results are the symbols, the lines guide the 
eye.  Dotted lines on left are assuming $\gamma_q=1$ all other results allow
for variation of $\gamma_q$ in the fit.  Dashed lines, on left, are for central 
rapidity, all other results are including full coverage of phase space. 
See Refs.~\cite{Letessier:2005qe,Rafelski:2004dp}
for further detail of model and method. 
  }
\vskip -.3cm
\end{figure*}

 As is seen in  the top left panel of
  Fig.\,\ref{PEST}, the rise of strangeness yield with 
centrality is faster than the rise of baryon number yield.
For the  most central head-on reactions, we  reach,   
at RHIC-200, $ s/B =9.6\pm 1$. In other words we make more than 
three strange quark pairs for every valance quark that is stopped
in the central rapidity region.    
In the middle  panel of \rf{PEST}   we compare strangeness with 
entropy production $s/S$. We see a 
smooth transition  as function of participant number $A$, (on left)
from a flat peripheral 
behavior where $s/S\lesssim 0.02  $  to smoothly increasing $s/S$
 reaching $s/S\simeq 0.028$
fro the  most central reactions.  On right, in \rf{PEST}, we see that 
the change on $s/S$ is much more drastic as function of reaction
energy. In both cases (energy and centrality dependence), the 
ratio  $s/S$  rises smoothly to the asymptotic value expected based
on the count of quark and gluons degrees of freedom inherent in \req{sdivS2}.

In the bottom panel of \rf{PEST}, on left, we see the thermal energy cost $E/s$ 
of producing a pair of strange quarks as function of the size of the participating 
volume ({\it i.e.\/}, $A$) This quantity shows a smooth 
drop which can be associated with transfer of thermal energy into collective
transverse expansion after strangeness is produced. Thus, it seems that the 
cost of strangeness production is independent of reaction centrality. The
result  is different when we consider $\sqrt{s_{\rm NN}}$ dependence of this 
quantity, see bottom panel on right. There is a very  clear change in the 
energy efficiency of making strangeness 
at the threshold energy.  

\section{DECONFINEMENT THRESHOLD AND PHASE STRUCTURE}
\label{compare}
Above results suggest that at, sufficiently high energy and large number
of participants, the system considered has as dynamical degrees of freedom
quarks and gluons. 
In order to strengthen the evidence for the new phase of matter, we would
like to be able to find experimentally and without 
a complex, and thus model dependent, analysis, some 
evidence for a change in reaction mechanism. We can consider  variation  
of particle yields   either with the centrality of collision, which 
fixes the initial transverse size of the fireball, and thus its size, or the 
variation with reaction energy. 
We are looking for a sharp change in particle yields. Since
these vary rapidly due to global changes in variables such as 
the total energy, and/or volume, the more interesting experimental
variable is the ratio of hadron yields.  

We can consider   a variable which traces out qualitatively 
 the observables we discussed in last section.
 Since the yield of K$^+$ closely follows 
that of strangeness $s=\bar s$, and the yield of $\pi^+$ is related to the 
total multiplicity $h$, and thus entropy $S$, the experimental
observable of interest is the ratio
K$^+/\pi^+ \propto \bar s/\bar d$ yield ratio~\cite{Glendenning:1984ta}. This 
ratio has been studied experimentally as function of $\sqrt{s_{\rm NN}}$~\cite{Gaz}  
and a pronounced `horn' structure  arises at relatively low reaction energies,
see  \rf{Kpisqrt}. Moreover there seems to be a raise in this ratio after a
dip at intermediate energies.

\begin{figure}[htp]
 \psfig{width=7.8cm,  figure=\pathnow 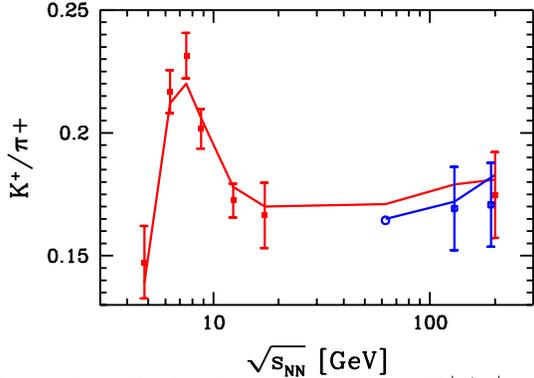}
\vskip -1.5cm
  \caption{\label{Kpisqrt}
Ratio of total yields of  $K^+/\pi^+$ as function of $\sqrt{s_{\rm NN}}$, with a fit   using 
SHARE and chemical non-equilibrium (red). At RHIC we also fit 
the central rapidity yields (blue). 
}\vskip -.6cm
\end{figure}

  This effect is  understood,  in data analysis, to be due to  
a rather sudden modification of   chemical conditions in the dense
matter fireball: the 
rapid rise in strangeness $\bar s$ production below,  and  a   rise in 
the anti-quark $\bar d$  yield above the peak. 
The measured ${\rm K}^+/{\pi}^+$ ratios are fitted well as is shown by the 
continuous line in  \rf{Kpisqrt}.  
 
The horn   arises solely in the one  $K^+/\pi^+$ particle ratio and in a good
approximation it   
traces out  the final state  valance quark ratio $\bar s/\bar d$. In language of
quark phase space occupancies $\gamma_i$ and fugacities $\lambda_i$, we have: 
\begin{eqnarray}
{{\rm K}^+\over \pi^+}&\to& {\bar s\over \bar d} =
   F(T)\left(\sqrt{\lambda_{I3}}{\lambda_s\over \lambda_q}\right)^{\!-1} 
                                   {\gamma_s\over \gamma_q}.
\end{eqnarray}

In chemical equilibrium models $\gamma_s/\gamma_q=1$, the isospin factor $\lambda_{I3}$ is insignificant.
Thus the horn in the  ${\rm K}^+/\pi^+$ ratio   
must arise solely from the variation in the ratio $\lambda_s/ \lambda_q$ and the 
change in temperature $T$. $F(T)$, which describes the ratio of phase spaces, 
is a smooth  function of $T$. Normally,
one expects that $T$ increases with collision energy, hence on this ground 
alone, we expect an
monotonic  increase  in the ${\rm K}^+/\pi^+$ ratio as function of reaction
energy.  

As the collision energy rises, the increased hadron yield per baryon requires 
  a decreasing value of $\lambda_q=e^{\mu_{\rm B}/3T}$. 
The two chemical fugacities  $\lambda_s$ and $\lambda_q$ are
coupled by the condition that the strangeness is conserved. This 
leads to a smooth  $\lambda_s/ \lambda_q$. Temperature of 
fit is usually   decreasing 
smoothly with increasing chemical potential. Thus, without the 
chemical non-equilibrium, one expects and finds  a smooth behavior of the
 K$^+/\pi^+$ ratio. Thus,  models which take chemical equilibrium as dogma fail
to describe this interesting experimental result.   

The  chemical freeze-out  conditions we have determined presents,    in the 
$T$--$\mu_{\rm B}$ plane, a more complex picture than naively expected,
 see     \rf{Tmu}. The RHIC $dN/dy$ results are to outer left.
They are followed by RHIC and SPS $N_{4\pi}$ results. The dip
corresponds to the 30 and 40 $A$GeV SPS results. The top right is 
the lowest  20 $A$GeV SPS and top 11.6 $A$GeV AGS energy range. 
To guide the eye, we have
added two lines connecting the fit results.  We see that the  
chemical freeze-out temperature 
$T$ rises for the two lowest reaction energies  11.6 and 20 $A$ GeV
 to near the Hagedorn temperature,  $T=160$ MeV, of boiling hadron  
matter. 

\begin{figure}[!tb]
\centerline{\psfig{width=8cm,figure=\pathnow     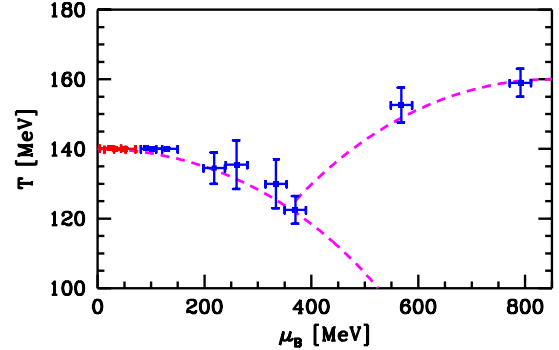}}
\vskip -1.3cm
\caption{\label{Tmu}
$T$--$\mu_{\rm B}$ plane with hadronization points obtained in 
the SHM fit to particle yields.  
}
\vskip -.8cm
\end{figure}

The shape of the  hadronization boundary,  shown in \rf{Tmu} 
in the $T$--$\mu_{\rm B}$ plane,   is the result
of a complex interplay between the dynamics of 
heavy ion reaction, and the properties of both phases of
matter, the inside of the fireball,  and the hadron phase we
observe.  The dynamical effect,  
capable to shift the location in temperature of 
the expected phase boundary is due to 
the expansion dynamics of the fireball see section \ref{exp},
and effects of chemical non-equilibrium, see section \ref{Tphase}
for full discussion. 
 
 Considering 
the sudden nature of the fireball breakup 
seen in several observables~\cite{RBRC}, 
we conjecture that the hadronizing fireball, leading to
 $\gamma_s>\gamma_q=1.6$ , super-cools and experiences  
 a true 1st order  phase transition also    at small 
$\mu_{\rm B}$. The system we observe 
in the final state prior to hadronization is mainly a quark--anti-quark
system with gluons frozen in prior expansion cooling of 
the QCD deconfined parton fluid. The quark dominance is necessary
to understand, {\it e.g.\/},  how  the azimuthal asymmetry $v_2$  
varies for different particles~\cite{Huang:2005nd}.

These quarks and anti-quarks have, in principle,  at that stage a significant 
 thermal mass. The evidence for this is derived from the
 dimensionless variable $E/TS$ (thermal energy divided by entropy and temperature). 
The energy  end entropy  per particle
 of  non-relativistic and semi-relativistic classical 
particle gas comprising both quarks and anti-quarks 
is  (see section 10, \cite{CUP}):
\begin{eqnarray}
{ E\over N}&\simeq& m+3/2\, T+\ldots,\\
{ S\over N}& \simeq& 5/2+m/T+\ldots,\\   
{E\over TS}&\simeq& {m/T+3/2\over m/T+5/2}.
\end{eqnarray}
The value  $E/TS\to 0.78$ found at large reaction energies and 
in most central collisions~\cite{Rafelski:2004dp,Letessier:2005qe} can be understood in 
 terms of a quark matter made of particles, with 
  $m\propto aT$, $a=2$  which is close
to what is  expected based on 
thermal QCD~\cite{Petreczky:2001yp}.

\section{LHC} \label{LHC}
We expect considerably more violent transverse expansion of the 
fireball of matter created at LHC, as compared to RHIC.  The kinetic energy of this
transverse motion must be taken from the thermal energy of the expanding matter.
This leads to a greater local cooling and thus to a greater reduction in the number of 
thermal quanta. The   entropy density decreases, but the expansion
assures that the total entropy remains constant, or slightly  increases. 

As the energy is transfered into the transverse expansion, it would 
appear that primarily  gluons are
feeding the expansion dynamics, while strange quark pair yield, being  weaker
coupled, remain least influenced by this dynamics.  
This mechanism  helps to   increase  the 
$K/\pi$ ratio as reaction energy increases, see  \rf{Kpisqrt}.
It is possible that   strange $s$, and $\bar s$ quark
densities can rival in magnitude  the 
light quark components and thus facilitate the phase transition. This should be seen 
in the   detail of the distribution of  particle yield. 

In this context,    more interesting than the $K/\pi$ ratio enhancement would be the 
enhancement anomaly in strange (anti-baryon) yields. With $\gamma_s\gg 1$, we  
 find that the more  strange  baryons and anti-baryons are more
abundant than the more `normal' species.   Specifically of interest would 
be $(\Omega^-+\overline\Omega^+)/(h^++h^-)$, 
$(\Xi^-+\overline\Xi^+)/(h^++h^-)$, and $2\phi/(h^++h^-)$ which 
show  an  order of magnitude shift in relative production
strength.  Detailed predictions for the yields of these particles
require considerable extrapolation of physics conditions from the RHIC to LHC
domain~\cite{LHCPred}.

We have so far not discussed charm over-saturation.  
Given the large charm quark mass, we  expect that
most of charm quark yield is due to first hard interactions of
primary partons. However, there is non-negligible thermal 
production {\it and} annihilation of charm pairs.  
The  yield of strange and light 
quarks, at time of hadronization, exceeds by about a factor 50   or more 
that of charm at central rapidity. Thus, even though charm 
phase space occupancy $\gamma_c$ at hadronization may largely exceed the 
chemical equilibrium value $\gamma_c\ggg 1$ given 
  hadronization temperature,   $m_c/T\simeq 10$--$30$, it 
takes a factor $\gamma_c\simeq e^{10}/ 10^{1.5}=700$ to compensate 
 particle  yield suppression due to the high charm mass. 

Said differently,
while strange quarks can compete in abundance  with light quarks considering
$m_s/T\simeq 1$, and $\gamma_s>1$, charm (and heavier) flavor(s) will remain suppressed,
in absolute yield, at the temperatures we can make presently in
 laboratory experiments. Consequently, they cannot play any significant
role in the dynamics of the phase transition.  

\section{HIGHLIGHTS}
Our objective has been to show that there is a good reason to 
expect that the behavior of the QGP formed in heavy ion collisions
can deviate in a significant manner from expectations formed in 
study of equilibrium thermal QCD matter. We have described two main
effects, the chemical non-equilibrium of strange quarks, and the 
pressure of flowing color charge acting on the vacuum, which, 
in our opinion, are at current level of knowledge relevant to the
issues considered. 
Of most theoretical relevance and interest are the 
implications of non-equilibrium hadronization on the possible
change in the location and {\it nature} of the 
 phase boundary.

We have further argued that strangeness, and entropy,
are  well developed  tools
allowing  the detailed  study of hot  QGP phase. A systematic study 
of strange hadrons  fingerprints the  properties 
of  a new state of matter at point of its breakup into final
state hadrons. 

We have shown that it is possible to describe the `horn' in 
the $K^+/\pi^+$ hadron ratio within the chemical non-equilibrium 
statistical hadronization model.  The appearance
of this structure is related to a rapid change in the properties 
of the hadronizing matter.


\end{document}